# Room temperature topological polariton laser in an organic lattice

M. Dusel, [1,*] S. Betzold, [1] T. H. Harder, [1] M. Emmerling,[1] J. Ohmer,[2] U. Fischer,[2] R. Thomale,[3] C. Schneider,[4] S. Höfling[1,5] and S. Klembt[1,#]

[1] Technische Physik, Wilhelm-Conrad-Röntgen-Research Center for Complex Material Systems, and Würzburg-Dresden Cluster of Excellence ct.qmat, University of Würzburg, Würzburg, Germany

[2] Department of Biochemistry, University of Würzburg, Würzburg, Germany

[3] Institute for Theoretical Physics and Astrophysics and Würzburg-Dresden Cluster of Excellence ct.qmat, University of Würzburg, Würzburg, Germany

[4] Institute of Physics, University of Oldenburg, D-26129 Oldenburg, Germany

[5] SUPA, School of Physics and Astronomy, University of St. Andrews, KY 16 9SS, United Kingdom

*marco.dusel@uni-wuerzburg.de

[#]sebastian.klembt@uni-wuerzburg.de

**Abstract**

**Interacting bosonic particles in artificial lattices have proven to be a powerful tool for the investigation of exotic phases of matter as well as phenomena resulting from non-trivial topology. Exciton-polaritons, bosonic quasi-particles of light and matter, have shown to combine the on-chip benefits of optical systems with strong interactions, inherited form their matter character. Technologically significant semiconductor platforms, however, strictly require cryogenic temperatures for operability. In this paper, we demonstrate exciton-polariton lasing for topological defects emerging form the imprinted lattice structure at room temperature. We utilize a monomeric red fluorescent protein derived from DsRed of Discosoma sea anemones, hosting highly stable Frenkel excitons. Using a patterned mirror cavity, we tune the lattice potential landscape of a linear Su-Schrieffer-Heeger chain to design topological defects at domain boundaries and at the edge. In spectroscopic experiments, we unequivocally demonstrate polariton lasing from these topological defects. This progress promises to be a paradigm shift, paving the road to interacting Boson many-body physics at ambient conditions.**

**Introduction**

Topology has emerged as an abstract, yet powerful paradigm for characterizing the flow of an excitation in a physical system, such as embodied by electrons or light. Since their systematic study and description by the 2016 Nobel Prize laureates Kosterlitz, Thouless [1], and Haldane [2], topological phenomena have soon been demonstrated in a range of physical systems. The field started off from two-dimensional electron gases [3,4] and spread to systems as diverse as cold atoms [5], mechanics [6], electric circuits [7], or microwaves [8]. Following visionary proposals to introduce concepts of topology into optical systems [9,10], topological photonics cover a wide range of research interests today, stretching from fundamental physics to future applications such as optical communication [11-13] and topological lasing [14-16]. In this context, exciton-polaritons resulting from the strong coupling of a quantum well exciton to a microcavity photon mode [17,18] have emerged as a powerful on-chip system for topological physics in hybrid-light matter systems [19-21]. The notable advantage of exciton-polaritons over purely photonic systems is the strong and tunable particle interaction inherited from the matter part of the hybrid system [22]. So far, the dominant experimental platforms are coupled micropillars etched into epitaxially grown GaAs-based microcavities [20,21,23-25]. While these structures provide superb material quality and mature technological control, the comparatively small exciton binding energy limits experiments to cryogenic temperatures. While exciton-polaritons have been studied in wide bandgap semiconductors like GaN or ZnO with some success [26,27], the overall material quality has so far inhibited the fabrication of controlled lattice potential landscapes. A new approach towards polaritonics at ambient conditions has been made using novel emitter materials, ranging from transition metal dichalchogenides [28-30] and perovskites [31,32] to a range of organic semiconductor materials [33-36]. The latter have the advantage that, due to their soft nature, they can easily be introduced into dielectric microcavities, including those with patterned potential landscapes [37,38].

In this work, we utilize mCherry, a member of the so-called mFruit family of monomeric red fluorescent proteins, hosting highly stable Frenkel excitons. The strong localization leads to high exciton binding energies defined by the difference between electronic and optical bandgap. The binding energy of a Frenkel exciton results from the on-site Coulomb interaction of electrons and hole and typically exceeds 100 meV [39]. By using a patterned mirror microcavity, we are able to control the lattice potential landscape with the precision of the focused ion beam used for the shaping. Here, we create a linear Su-Schrieffer-Heeger chain hosting (intentionally created) topological defects at a domain boundary and at the edge of the chain. These defects are located in a sizeable topological gap of ~4.9 meV that exceeds the linewidth of the topological modes which is of the order of 385 µeV in the linear regime. Using advanced spectroscopic techniques including tomography methods we demonstrate clear evidence for polariton lasing from topological domain boundary and edge defects at room temperature.

**Results**

The device used in this study is composed of two distributed Bragg reflectors enclosing a thin layer of mCherry. By constructing a lattice of hemispheric structures in one of the mirrors, an array of coupled microcavity sites is created. The Su-Schrieffer-Heeger model is derived for the description of alternating double and single carbon bonds in the polymer polyacetylen and has a unit cell comprised of two sites [40]. The Hamiltonian in the tight binding limit is given by

$$\hat{H} = v \sum_{m=1}^{N} (|m,B\rangle\langle m,A| + h.c.) + w \sum_{m=1}^{N-1} (|m+1,A\rangle\langle m,B| + h.c.),$$

where $N$ denotes the number of unit cells, $A$ and $B$ the sites in the unit cell, and $v$ and $w$ the respective intra- and inter-cell hopping amplitudes [41]. From this Hamiltonian two topologically distinct phases for the cases (i) $v < w$ and (ii) $v > w$ emerge. To show that these cases are topologically distinct, we calculate the difference in phase winding

$$W = \frac{1}{2\pi} \int_{BZ} \frac{\partial \varphi(k)}{\partial k} dk,$$

where the winding $W$ of the phase $\varphi(k)$ is calculated across the Brillouin zone. As a result, a weakly bound edge pillar or a domain boundary defect having weak couplings in all directions ($v < w$) lead to a topological defect and a bulk winding number of $W = 1$, whereas the opposite case of ($v > w$) of strongly coupled defects remains topologically trivial and the winding number is $W = 0$. Contrary to previous realizations of the SSH model for polaritons in zigzag chains relying on the orientation and coupling of $P_{x,y}$-modes [20,42,43], we use a linear version for the lowest energy $S$-mode. Thus, the weak and strong inter-site couplings are not achieved by an overlap asymmetry of higher order modes, but by sheer difference in distance and overlap for circular-symmetric modes.

In Fig. 1 (a), an artistic rendering of our device is depicted. A thin layer of the red fluorescent protein mCherry (see inset) is embedded between two DBRs, consisting of 10 alternating pairs of SiO$_2$ and TiO$_2$, where one of the mirrors is patterned to invoke the lattice potential landscape. The chain is comprised of a total of 50 sites (Supplementary Section 1). The hemispheric lattice sites have a diameter of 3 µm and a curvature radius of 13 µm. The layer of mCherry has an approximate thickness of 1.5 µm. To implement the SSH model, a weaker coupling is realized by a larger site-to-site distance of 1.95 µm and a the stronger coupling is realized by a distance of 1.56 µm. The inset depicts the molecular structure of mCherry, with the chromophore (purple) surrounded by an 11-stranded β-barrel. Fig. 1 (b) shows the absorption and emission spectra of a typical layer of mCherry. The material features a significant Stokes shift between the broad absorption maximum at $E_{abs} = 2.11$ eV and the emission maximum at $E_{em} = 1.98$ eV that is common for organic semiconductors hosting Frenkel excitons. Fig. 1 (c) depicts a schematic of the couplings, with the topological defects marked in red at the edge of the chain and at a domain boundary, where the pattern of weak coupling following strong coupling is broken by two consecutive weak couplings. The defects are well separated by 15 dimers between them allowing us to investigate them independently. The microcavity arrangement yields strong light-matter coupling with an excitonic transition at $E_X = 2.085$ eV, a Rabi splitting $2\hbar\Omega_R = 215$ meV and an exciton-photon detuning of $\delta = -230$ meV, yielding an excitonic fraction of the polariton of ~10% (Supplementary Section 2).

All spectroscopic measurements presented in this paper were performed at room temperature, specifically at $T = 293$ K. The sample is mounted on a *(x,y)*-piezo stage in a hanging configuration and excited with a continuous wave laser at $\lambda_L = 532$ nm resonant with the first high-energy Bragg minimum of the top DBR mirror. Fig. 2 shows a spectroscopic characterization of the device. In Fig. 2 (a) the elliptical laser spot is centered on the domain boundary SSH defect, while in (b) the laser is placed on the defect-free bulk of the chain. The angle-resolved photoluminescence spectra reveal a distinct and sizeable gap of ~4.9 meV resulting from and scaling with the coupling difference $\Delta = |w - v|$. In addition to the gap, Fig. 2 (a) displays a distinct mode within the topological gap, originating from the topological SSH domain boundary defect. The linewidth of the defect mode is $\gamma_{topo} = 385$ µeV, well within the topological gap. In Fig. 2 (b) there is no emission in the gap as expected from a defect-free system. The dispersion band structure of the gapped *S*-band is accurately reproduced using a tight-binding model.

To further investigate the nature of the defect, we now similarly excite the domain boundary defect and plot the emitted energy below any non-linear threshold along the chain in space in Fig. 2 (c). The anti-binding *S*-band above the topological gap at $E_S^{anti} = 1.8220$ eV highlighted as a mode tomography in Fig. 2 (e) is revealing the exact position of each site, as expected, while the binding *S*-band below the gap at $E_S^{bin} = 1.8132$ eV in Fig. 2 (g) is underlining the dimerized nature of the chain. Note that the intensity is magnified as compared to the defect mode. However, the photoluminescence signal is of a similar intensity for at least 12 lattice sites shown. Contrary to that, the defect mode is exponentially localized around the defect site. In fact, Fig. 2 (f) shows the main intensity centered on the defect, and in addition the typical signature of a topological SSH defect, where exponentially decaying emission intensity from one site of each unit cell is visible. To highlight the defect localization and distinct nature of the SSH defect, Fig. 2 (d) shows the intensity of the anti-binding *S*-band and the defect, normalized to the maximum intensity, respectively. Here, the exponentially localized defects mode reveals exactly the mode pattern expected from a topological 'weak-weak' (or 'long-long') defect in an SSH chain [44].

Having established the typical features of a topological SSH defect, we use non-resonant pulsed laser excitation at $P_{L,cw} = 532$ nm to excite the same chain with the laser spot centered at the domain boundary defect characterized in Fig. 2. Most studies devoted to polariton condensation using organic emitter materials pump with short laser pulses (<10 ps) to circumvent exciton−exciton annihilation at high power densities. The molecular structure of mCherry however can sustain extended laser pulses of 6 ns at a repetition rate of 2 Hz like in this study. The reason is that the chromophore is enclosed by a so-called β-barrel, protecting the chromophore from the environment. Furthermore, the barrel structure effectively reduces exciton−exciton annihilation and suppresses biomolecular quenching at high excitation densities [36,37]. A respective angle-resolved PL spectrum is shown in Fig. 3 (a). The dispersion at an excitation power of $P{\sim}0.08$ nJ/pulse is qualitatively similar to the one employing continuous wave excitation. Note however, that the overall low intensity results for the low pulse repetition rate of 2 Hz and thus the weak total signal. The excitation power is subsequently increased to (b) $P{\sim}0.10$ nJ/pulse, and (c) $P{\sim}0.21$ nJ/pulse. Fig. 3 (d) shows the integrated photoluminescence intensity and linewidth of the topological domain boundary defect as a function of the laser excitation power. At the polariton condensation threshold $P_{thr}^{dom}{\sim}\,0.1$ nJ/pulse a sudden decrease of the linewidth towards the resolution limit of the spectrometer, marking the build-up of phase coherence, accompanied by a strong non-linear increase of the output intensity by approximately three orders of magnitude is observed. When following the topological defect energy across the condensation threshold an interaction-induced blueshift of ~1.5 meV is observed across the variation of excitation powers (Fig. 3 (e)). In conventional GaAs-based cavities, polariton-polariton interactions and interactions with the exciton reservoir mainly cause inter-particle repulsion. Previous studies have shown that, in contrast, the density-dependent blue shift in Frenkel excitons arises largely due to the screening of Rabi splitting [37,45]. Furthermore, it has been shown that the underlying mechanism causing the interaction is not relevant regarding the phenomenological observations in the nonlinear regime [38]. Importantly, the defect always remains well within the topological gap, even when assuming near-zero blueshift of the bulk modes [20,43]. It is important to note, that the linear configuration used in this work easily allows for the realization of a domain boundary defect, while in the typically employed polariton zigzag SSH chains [20,42,43] so far only edge defect could be studied. Since the linear chain easily allows for both defect types, we have performed the similar non-resonant pulsed laser excitation experiment on the topological edge defect, characterized by a weak terminal bond. To highlight the geometry of this defect, Fig. 4 (a-c) show the mode tomographies of (a) the antibinding *S*-band, (b) the topological edge defect and (c) the binding *S*-band. Similarly, to the domain boundary defect the edge defect is exponentially localized, revealing the distinct decrease of PL signal on every second site (Fig. 4 (b)). Fig. 4 (d)-(f) display the angle-resolved dispersion measurements for increasing laser excitation power. The linewidth and output power plotted in Fig. 4 (g) show the distinct occurrence of a polariton condensation threshold at $P_{thr}^{edge}$ ~ 0.25 pJ/pulse and a continuous blueshift of the mode of 0.4 meV in total (Fig. 4 (h)). We find that the exact value of excitation power at which the threshold occurs delicately depends on the positioning of the excitation laser on the chain, thus being the likely explanation for the deviation in threshold powers between the two defect types.

The very nature of exciton-polaritons is that of a driven dissipative system, naturally involving gain and loss. For this reason, photonic systems in general and polariton systems in particular are well suited for the study of non-Hermitian photonics, possibly enriched by broken or unbroken parity-time symmetry PT. Combined with the realizability of different topological excitations such as boundary or domain wall defect modes [46], exciton-polaritons open new perspectives of topological design at room temperature.

**Conclusion**

In conclusion, we have presented a versatile and comparatively simple microcavity platform using a fluorescent protein allowing for stable excitons, and thus exciton-polaritons, at room temperature. In this coupled cavity device we have realized polariton lasing from topological SSH defect modes both at a domain boundary and at the edge. The polariton condensation is robustly occurring in the defects, and the modes remain well within the topological gap despite an interaction-induced blueshift. Our organic microcavity platform holds great potential to emulate interacting boson many-body physics and lattice emulation on chip and at room temperature. In addition, given the driven-dissipative nature of the system, our platform promises unprecedented opportunities for studies of PT-symmetry in a hybrid light-matter system [47] by accurately controlling on-site gain and loss, for example by using spatial light modulators.

**Acknowledgements**

M.D., S.B., M.E., R.T., S.H., and S.K. acknowledge financial support by the German Research Foundation (DFG) under Germany's Excellence Strategy–EXC2147 "ct.qmat" (project id 390858490). S.K. acknowledges support by the German Research Foundation (DFG) within project KL2431/2-1. S.H. is furthermore grateful for support within the EPSRC Hybrid Polaritonics Grant (Grant No. EP/M025330/1). T.H.H. and S.H. acknowledge funding by the doctoral training program Elitenetzwerk Bayern Graduate School "Topological insulators". T.H.H. acknowledges support by the German Academic Scholarship Foundation.

**Methods**

**Sample preparation.** Our microcavity device consists of two distributed Bragg reflectors enclosing a thin layer of the fluorescent protein mCherry. The in-plane confinement and SSH lattice structure is realized by patterning one of the mirrors while the other is planar. The patterned mirror is created using ion beam milling (FEI Helios NanoLab DualBeam) to create dimples arranged in a linear SSH chain on a quartz substrate with an roughness of <1 nm RMS. The hemispheric dimples have a diameter of 3 µm and depth of the order of 175 nm. The site spacing is 1.95 µm for the short coupling distance and 1.56 µm for the short distance. The quartz sample with the milled dimples is subsequently covered with a DBR using ion beam sputter deposition (Nordiko 3000) of 10.5 pairs of alternating $SiO_2$ (103 nm) / $TiO_2$ (65 nm). A highly concentrated solution (175g/Liter) of mCherry was spin-coated on the patterned DBR. The device is completed by placing a planar $SiO_2/TiO_2$ DBR on the mCherry. Finally, the device is dried out for 48 h under a constant pressure of ~0.25 N/cm2. The structure yields an experimental microcavity Q-factor of 4720 and a Rabi splitting of ~215 meV.

**Sample characterization.** For basic characterization of the samples in the linear regime we use a 532 nm continuous-wave diode laser that is resonant with the first high-energy Bragg mode of the top mirror. For the nonlinear regime we use a wavelength-tunable optical parametric oscillator system with nanosecond pulses tuned to 532 nm. The laser beams where focused onto the sample surface using an elliptical spot of 5 μm ∗ 20 μm or a spot size diameter of ~5 μm using a microscope objective with a numerical aperture of 0.42. Photoluminescence emission was collected in reflection geometry using the same objective. The signal was filtered using a 550 nm long-pass filter brought to the entrance slit of a 500 mm Cherny-Turner spectrometer and recorded by a Peltier-cooled electron multiplying charge-coupled device (Andor Newton 971) with a chip size of $1600x400$ pixel and a pixel size of $16x16$ μm² . For energies around 1.8 eV the spectral resolution of the system is ~150 μeV. The dispersion measurements were performed using a Fourier imaging configuration collecting the angle-dependent (momentum space) signal for the back-focal plane of the microscope objective. The mode tomographies were created using a motorized imaging lens which moves the real-space image across the spectrometer slit and thus the full information in $(x, y)$-space and energy can be retrieved.

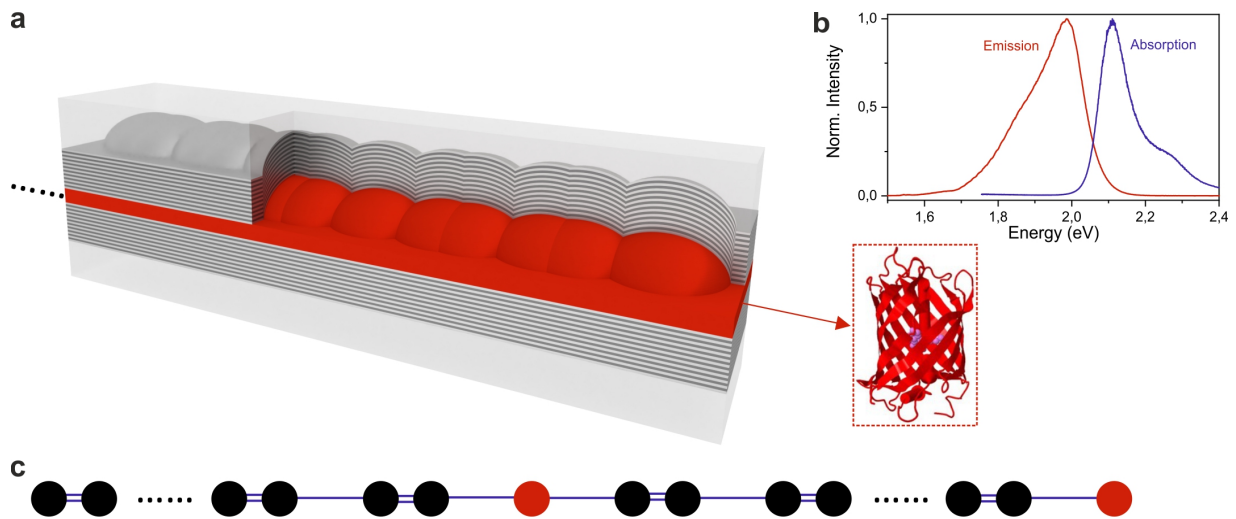

**FIG. 1 (a)** Artistic illustration of a linear, one-dimensional Su-Schrieffer-Heeger chain of microcavities containing a thin film of mCherry. The individual sites created by hemispheric mirrors that are arranged such that they ensure alternating weak and strong inter-site coupling. Inset: Schematic molecular structure of mCherry, with the chromophore (purple) surrounded by an 11-stranded b-barrel. (b) Absorption (blue) and emission spectrum (red) of a mCherry film with a dominant absorption at 2.11 eV and dominant emission at 1.98 eV, displaying a significant Stokes shift. (c) Schematic representation of the couplings in a linear Su-Schrieffer-Heeger chain. A topological edge defect as well as a domain boundary defect (red circles) are incorporated.

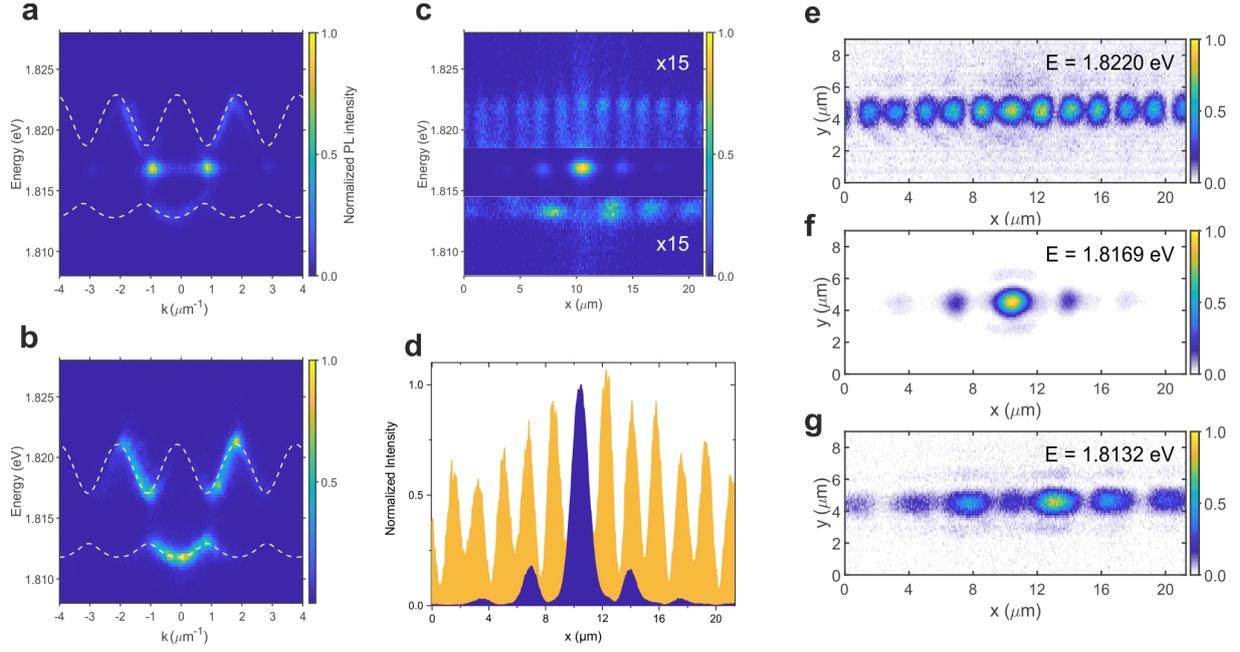

**FIG. 2** (a) Angle-resolved photoluminescence spectrum with the laser excitation spot centered on the domain boundary defect, and (b) on the bulk of the chain. In both cases, the spectrum shows a distinct gap of 4.9 meV in the lowest energy band (*S*-band) between 1.8144 eV and 1.8193 eV. When exciting the defect in (a) emission at $E_{topo}^{dom} \sim 1.8169$ eV with a linewidth of 385 μeV is visible, while the bulk chain in (b) reveals no emission from the gap as expected. (c-g) Respective real-space spectra of the SSH chain including the domain boundary defect (c). While the anti-binding (e) and anti-binding band (g) in the mode tomographies reveal the individual lattice sites, the mode within the gap in (c) shows the distinct signature on an SSH defect (f). (d) Line scan of the normalized intensity of the anti-binding S-mode (yellow) and the defect (blue) highlighting the exponential localization of the defect.

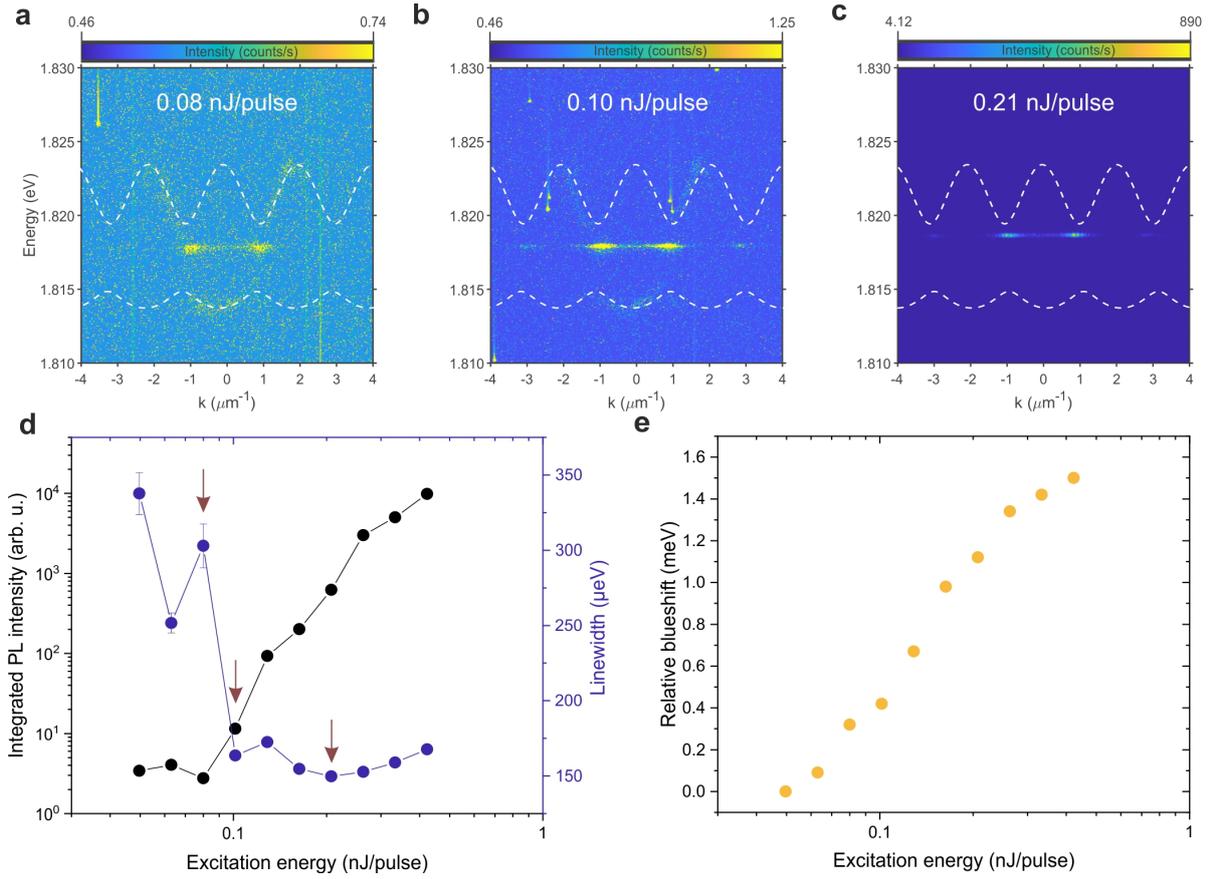

**FIG. 3** Angle-resolved photoluminescence spectra from the domain boundary defect for increasing laser excitation power (a) $P{\sim}0.08$ nJ/pulse, (b) $P{\sim}0.10$ nJ/pulse, and (c) $P{\sim}0.21$ nJ/pulse. (d) Integrated photoluminescence intensity and linewidth of the topological domain boundary defect as a function of the laser excitation power. At the polariton condensation threshold $P_{thr}{\sim}0.10$ nJ/pulse a sudden decrease of the linewidth towards the resolution limit of the spectrometer accompanied by a strong non-linear increase of the output intensity by approximately three orders of magnitude is observed. The three arrows indicate the dispersions plotted in (a-c), respectively. (e) Evolution of the emission energy as a function of the laser excitation power. A continuous increase of an interaction-induced blueshift with a total shift of ${\sim}1.5$ meV caused by the exciton-polariton nature of the system is observed.

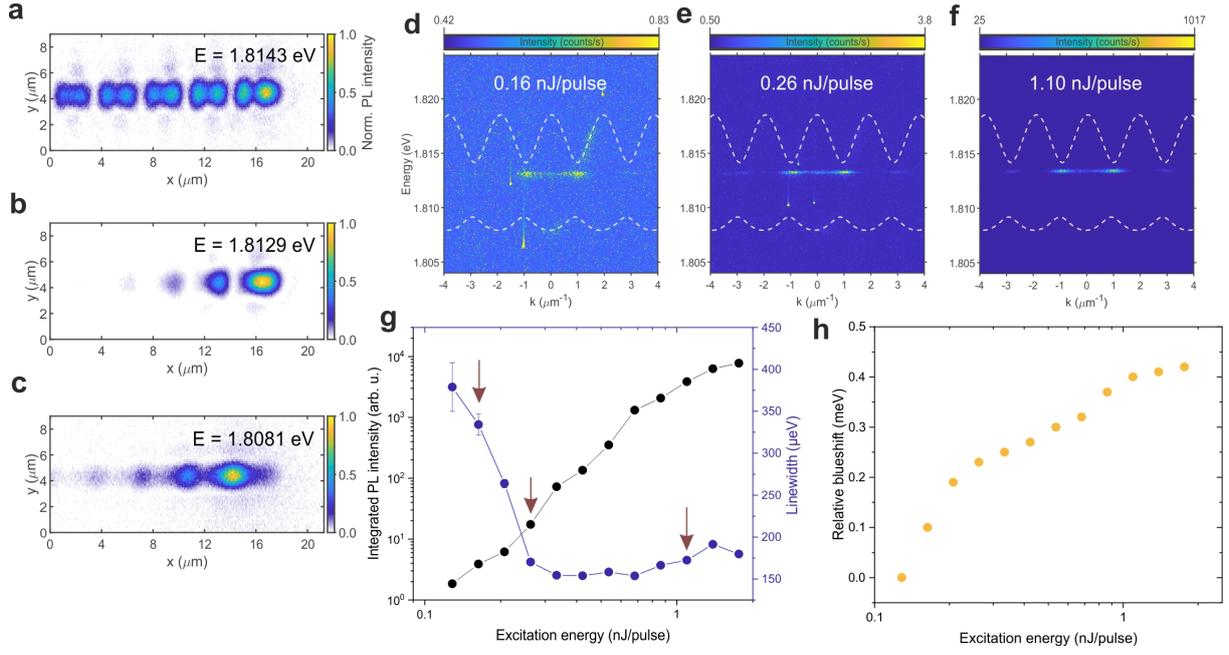

**FIG. 4** (a-c) Real-space mode tomographies of the (a) anti-binding *S*-band ($E_S^{anti} = 1.8143$ meV), (b) the edge defect ($E_{topo}^{edge} = 1.8129$ meV) and (c) binding *S*-band ($E_S^{bind} = 1.8081$ meV) below threshold. (d-f) Angle-resolved photoluminescence spectra from the edge defect for increasing laser excitation power (d) $P\sim0.16$ nJ/pulse, (e) $P\sim0.26$ nJ/pulse, and (f) $P\sim1.10$ nJ/pulse. (g) Integrated photoluminescence intensity and linewidth of the topological edge defect as a function of the laser excitation power. Here, the polariton condensation threshold occurs at $P_{thr}\sim0.25$ nJ/pulse revealing the typical behavior of a polariton laser. (h) Evolution of the emission energy as a function of the laser excitation power. A continuous increase of an interaction induced blueshift totaling $\sim0.4$ meV caused by the exciton-polariton nature of the system is observed.

# Supplementary information

**S1 Microscopy image of patterned mirror**

Suppl. Fig. 1 shows a microscope image of the patterned DBR mirror prior to the spin-coating of mCherry. All SSH chains display a total of 50 sites. The domain boundary defects is marked with "X" while the edge defect is the lowermost site (red circles). The varying chain lengths are a result of the variation of the coupling lengths reduced spacing $v = a/d$ with the center-to-center distance $a$ normalized by the diameter $d$. The SSH chain and the two defects presented in this work is marked (red circles; black dashed line). For this chain the weaker and stronger couplings are linked to $v_1 = 0.65$ and $v_2 = 0.52$.

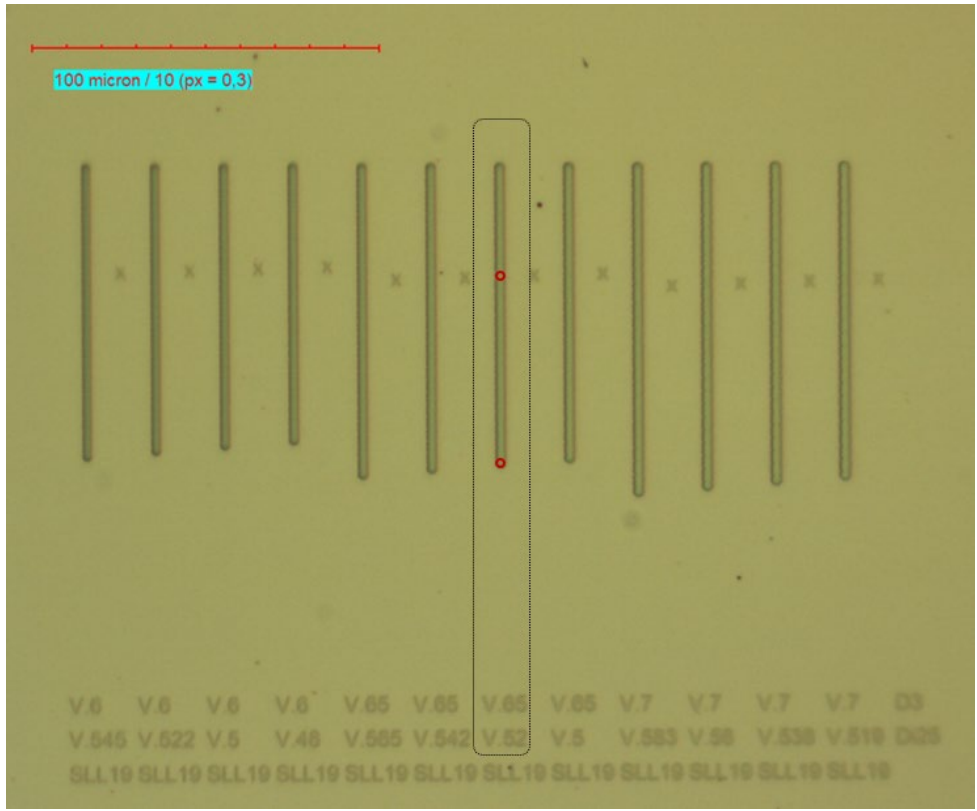

Suppl. Fig. 1: Microscopy image of the patterned DBR mirror prior to the spin-coating with mCherry. Chains with a total of 50 sites and varying couplings are show, with the reduced spacings $v$. The marked SSH chain was used in this work.

## S2 Strong light-matter coupling

An angle-resolved photoluminescence spectrum from the planar microcavity is used to study the strong light-matter coupling. Due to the comparatively large cavity length the excitonic transition at $E_X = 2.085$ eV couples to a number of photonic cavity modes ($C1 - C3$). A coupled oscillator model yields a Rabi splitting of $2\hbar\Omega = 215$ meV in good agreement with previous studies [49,50]. The resulting upper and lower polariton branches (UP, LP1, LP2) accurately reproduce the dispersion data. For the spectroscopic measurements in this work in the energy range of 1.81 eV this yields a excitonic fraction of the polariton of ~10%.

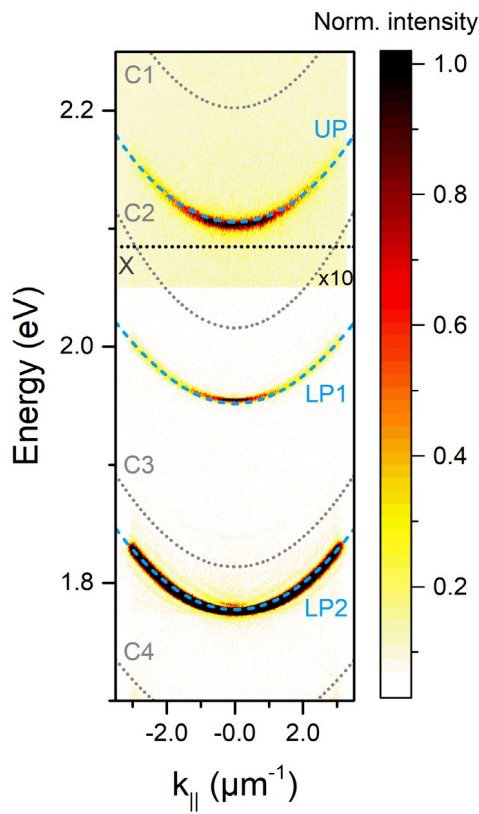

Suppl. Fig. 2: Angle-resolved photoluminescence spectrum from the planar microcavity area. A coupled oscillator model yields an exciton energy of 2.085 eV and a Rabi splitting of 215 meV.